\documentclass[aps,prx,twocolumn,amsmath,amssymb]{revtex4-1}
\usepackage{graphicx}
\usepackage[version=4]{mhchem}
\usepackage{bm}
\usepackage{braket}
\usepackage{natbib}
\usepackage{xcolor}
\usepackage{algorithm}
\usepackage{algpseudocode}
\usepackage{enumitem}
\usepackage{tikz}
\usetikzlibrary{arrows.meta,positioning,calc}

\definecolor{c1}{HTML}{FDE725}   
\definecolor{c2}{HTML}{440154}   
\definecolor{c3}{HTML}{414487}   
\definecolor{c4}{HTML}{7AD151}   
\definecolor{c5}{HTML}{22A884}   

\begin{document}

\title{Electron–molecule scattering via R-matrix variational algorithms on a quantum computer}

\author{Dario Picozzi}
\email{picozzi.dario@gmail.com}
\affiliation{Department of Physics and Astronomy, University College London (UCL), Gower Street, London, WC1E 6BT, United Kingdom}
\affiliation{London Centre for Nanotechnology, 19 Gordon St, London, WC1H 0AH, United Kingdom}

\author{Jonathan Tennyson}
\affiliation{Department of Physics and Astronomy, University College London (UCL), Gower Street, London, WC1E 6BT, United Kingdom}

\author{Vincent Graves}
\affiliation{School of Physical Sciences, The Open University, Walton Hall, Milton Keynes MK7 6AA, United Kingdom}
\affiliation{National Quantum Computing Centre, Rutherford Appleton Laboratory, Didcot OX11 0FA, United Kingdom}

\author{Jimena D. Gorfinkiel}
\affiliation{School of Physical Sciences, The Open University, Walton Hall, Milton Keynes MK7 6AA, United Kingdom}

\begin{abstract}
Electron--molecule collisions play a central role in both natural processes and modern technological applications, particularly in plasma processing. Conventional computational strategies such as the R-matrix method have been widely adopted yet encounter significant scaling challenges in treating more complex systems. In this work we present a quantum computational approach that utilises the variational quantum eigensolver (VQE) and variations thereof to overcome these limitations. We explore a number of methods, including the use of number projection operators and simultaneous optimisation. We demonstrate the feasibility of our method on a model problem of electron scattering from the hydrogen molecule, with numerical results obtained using a noiseless classical simulator. We recover the full spectrum of the Hamiltonian within a chosen symmetry sector. Moreover, the optimal circuit parameters directly encode the R-matrix boundary amplitudes needed for subsequent scattering computations. To our knowledge, this is the first application of quantum algorithms to electron--molecule scattering, and specifically the first formulation of the R-matrix inner-region problem on a quantum computer.

\end{abstract}

\maketitle

\section{Introduction}
Electron--molecule collisions drive a myriad of natural processes and are instrumental in modern technologies, especially in plasma processing \cite{Schneider1999, Carr2012, Zammit2016}. Studies of these processes increasingly rely on quantum mechanical calculations rather than on experiments \cite{16BaKu}. However, for many problems of interest the computational resources needed for highly accurate solutions scales so drastically that approximate methods are routinely employed. 

A host of quantum algorithms have been developed over the years. The idea of simulating quantum systems on a computer originated with Feynman \cite{Feynman1982} and was later formalised in Lloyd’s seminal work on universal quantum simulators \cite{Lloyd1996}. Early proposals for quantum phase estimation applied to chemistry were developed by Abrams and Lloyd \cite{Abrams1997, Abrams1999} and led to the first full simulation of molecular energies on a quantum computer by Aspuru-Guzik \textit{et al.} \cite{Aspuru2005}. More recently, in the noisy intermediate-scale quantum (NISQ) era, the variational quantum eigensolver (VQE) \cite{Peruzzo2014} – later refined by McClean \textit{et al.} \cite{McClean2016} – has become a popular method for electronic structure calculations. Early experimental demonstrations include implementations using photonic platforms \cite{Lanyon2010} and superconducting qubits \cite{OMalley2016}; further experiments on trapped ions have also been reported \cite{Hempel2018}. Reviews covering both methodology and applications now abound in the literature \cite{Cao2019, McArdle2020, Bauer2020, Tilly2022, Bharti2022}. Despite their promise, variational quantum algorithms (VQAs) can face significant optimisation challenges. More recent hardware demonstrations have scaled and improved VQE-based quantum chemistry implementations, including optimized unitary coupled-cluster approaches \cite{Guo2024NatPhysUCC} and trapped-ion demonstrations targeting relativistic molecular properties \cite{Chawla2025PRAEDM}.
In particular, the training landscape may exhibit {barren plateaus}, where gradients concentrate and become exponentially small with system size for broad classes of ans\"atze and cost functions, severely hindering trainability \cite{McClean2018BarrenPlateaus}.
Relatedly, recent work has highlighted that circuit structures which {provably} avoid barren plateaus can, in some settings, also imply restrictions that enable efficient classical simulation of the loss (or parts of it), sharpening the need to understand when trainability translates into genuine quantum advantage \cite{Cerezo2025AbsenceBP}.
These considerations motivate the use of problem-structured, symmetry-preserving, and relatively shallow circuit constructions, and cost functions that remain close to physically meaningful objectives (here, sequential energy minimisation within a fixed symmetry sector).

In parallel, quantum algorithms for scattering have been investigated. Jordan, Lee and Preskill proposed quantum algorithms for calculating scattering amplitudes in quantum field theories \cite{Jordan2011, Jordan2012, Jordan2014}, while recent proposals \cite{Choi2021, Klco2016, Arute2020, Yusf2025} demonstrate the progress toward simulating scattering on quantum devices. 

While quantum algorithms for scattering have been explored in a variety of contexts, prior work has not addressed electron--molecule collisions. For example, Xing \textit{et al.} used a variational quantum linear solver within the Kohn variational framework to compute multichannel $S$-matrices for {atom--molecule} collisions \cite{Xing2023_KohnVQLS}, and grid-based first-quantized simulations have treated electron scattering and ionization primarily in atomic model systems \cite{Chan2023}. Other demonstrations on quantum hardware have extracted two-body scattering phase shifts in effective field theory or gauge-theory settings \cite{Sharma2024,Yusf2025}. By contrast, here we formulate and demonstrate a quantum algorithm for the {electron--molecule} case, mapping the R-matrix inner-region problem to qubits and variational circuits. To our knowledge, this constitutes the first application of quantum algorithms to electron--molecule scattering and the first quantum-computational realization of the R-matrix inner region.

More broadly, this is, to our knowledge, first application of a Clebsch--Gordan (CG) symmetry–enforcing circuit to quantum scattering problems: we enforce a fixed spin symmetry through a two-qubit Givens block in a variational scattering ansatz and read out scattering boundary amplitudes directly from the optimised angles, a design not used in prior QC scattering demonstrations~\cite{Sharma2024,Xing2023_KohnVQLS}.

Simultaneous optimisation of multiple eigenstates has been explored in the subspace-search variational quantum eigensolver (SSVQE)~\cite{Nakanishi2019} and multistate contracted variational quantum eigensolver (MCVQE)~\cite{Parrish2019}. Our approach differs in three fundamental ways.
(i) {Sequential subspace optimisation (SSO):} SSVQE/MCVQE keep a batch of states orthonormal under a shared unitary but still require measuring the subspace Hamiltonian (off-diagonal $H_{ij}$) and a classical diagonalisation to extract individual eigenstates. In contrast, our fixed-order $SO(k)$ cascade block-diagonalises $H$ one state at a time, so each eigenstate is obtained directly by minimising only $\langle H\rangle$, with orthogonality enforced by circuit structure, without the need for penalty terms, overlap measurements, or subspace diagonalisation.

This structure lets us avoid squared ($H^2$) terms and pair with a coherent-summation readout over the selector register to cut the number of distinct expectation values. The optimised $SO(k)$ network determines the open-channel coefficients that enter the R-matrix boundary amplitudes, giving problem-native outputs beyond generic multistate VQE.

In this work we:
\begin{enumerate}
    \item Present a quantum‐algorithm framework for the R‐matrix inner‐region problem, mapping the $(N+1)$‐electron Hamiltonian to qubits via a Jordan–Wigner transformation and explicitly enforcing occupancy constraints for continuum orbitals.
    \item Employ the qubit representation of the number‐projection operator in the variational circuit—based on the construction in \cite{picozzi2024pascal}—to ensure that at most one electron occupies any continuum orbital.
    \item Develop a {sequential subspace optimisation} (SSO) protocol that decomposes the required $\mathrm{SO}(k)$ rotations into a fixed cascade of two-qubit Givens rotations (i.e.\ rotations that mix the $\ket{01}$ and $\ket{10}$ subspace), block-diagonalising $H$ one eigenstate at a time and enforcing orthogonality by construction, without the need to introduce penalty terms, overlap measurements, or subspace diagonalisation.
    \item Introduce a coherent‐summation measurement scheme for the target‐state selector register, which combines ancilla‐based Hadamard gates with system‐Hamiltonian measurements to reduce the total number of distinct expectation values required.
    \item Demonstrate our approach on a model $\mathrm{H}_2$ scattering calculation, validating the algorithm and showing that the optimal circuit parameters determine the R‐matrix boundary amplitudes needed for subsequent scattering analysis. The numerical results reported were obtained on a noiseless classical simulator (i.e., no device noise is included).  Unless stated otherwise, expectation values are evaluated exactly (statevector expectation values) rather than by finite-shot sampling, so there is no statistical measurement uncertainty to report as error bars.
\end{enumerate}

\section{The R-matrix method and molecular Hamiltonian}

The R-matrix method was originally developed to study nuclear \cite{Wigner1947}, and then atomic and molecular collisions \cite{Wigner1947,Burke2011}.  It is now \cite{jt474}  in widespread use to study electron-collisions for a whole range of molecular
problems and is also employed to study photoionisation and molecules in strong fields \cite{Masin2020}. The formulation of the R-matrix method is to divide the problem into a scattering-energy-independent inner region which, for electron -- molecule collisions, has strong similarities with problems of quantum chemistry \cite{jt474}, and an energy-dependent outer region. Complicated outer region problems can be computationally expensive but are well-suited for modern, massively parallel computations \cite{Sunder98}. Conversely the inner region, which is generally treated using variational methods, shows scaling  similar to that encountered
in quantum chemistry electronic structure problem but with extra complications which are discussed below.

In this paper we propose and provide formulations for solving the R-matrix inner region problem using an extension of the VQE framework. Two of us developed an R-matrix with pseudostates (RMPS) procedure for studying electron-molecule scattering \cite{jt341,jt354} which has shown to provide robust numerical solutions to challenging scattering problems \cite{jt434,jt510}. However,
the computational demands of this RMPS procedure scale so sharply  with the number of electrons that need to be
considered, that it remains largely unused despite ever-increasing demands for accurate cross sections.
A quantum computational implementation of the R-matrix method therefore has very significant potential uses and benefits.

\begin{figure*}[t]
  \centering
  \includegraphics[width=0.6\textwidth]{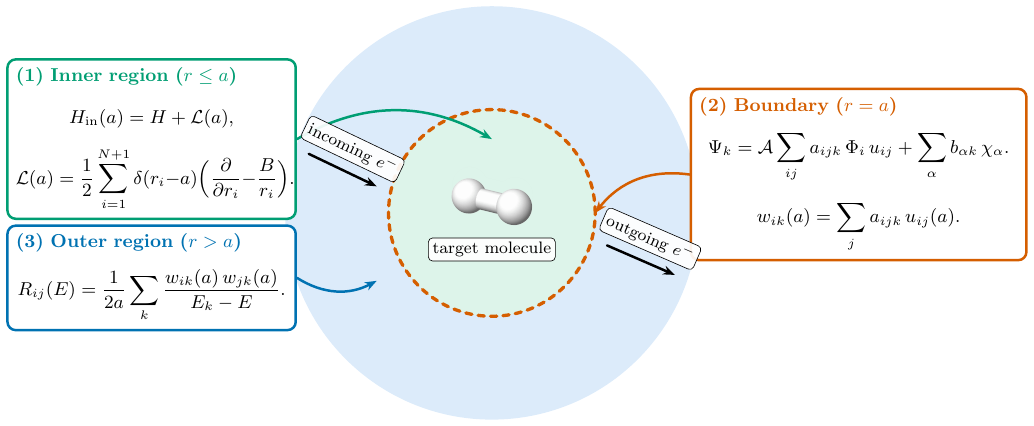}
  \caption{Schematic of the R-matrix partition for electron--molecule scattering. The inner region is the sphere of radius $a$ containing the molecular target, where the Hermitian inner-region operator $H_{\mathrm{in}}(a)=H+\mathcal{L}(a)$ is solved. At the boundary $r=a$, the inner-region eigenstates define the surface amplitudes $w_{ik}(a)$ entering the R-matrix $R_{ij}(E)$, which provides the matching condition for propagation in the outer region.}
  \label{fig:rmatrix_partition}
\end{figure*}

The R‐matrix formalism addresses scattering problems by partitioning space into an inner region and an outer region \cite{jt474} (Fig.~\ref{fig:rmatrix_partition}). In the inner region, the full many–electron Hamiltonian is solved within a confined volume that encompasses the molecular target, allowing for a detailed treatment of electron correlation, exchange, and short–range interactions.  

In our work we use the standard second–quantised representation of the molecular Hamiltonian, which is given by
\begin{equation}
H = \sum_{pq} h_{pq} a^{\dagger}_p a_q + \frac{1}{2} \sum_{pqrs} h_{pqrs} a^{\dagger}_p a^{\dagger}_q a_r a_s + h_{\mathrm{nuc}},
\label{eq:second_quantized_Hamiltonian}
\end{equation}
where $h_{pq}$ are the one–electron integrals (accounting for kinetic energy and electron–nucleus attraction), $h_{pqrs}$ are the two–electron integrals (accounting for electron–electron repulsion), and $h_{\mathrm{nuc}}$ represents the nuclear repulsion energy. In the inner region these integrals are computed over a finite volume (see below).

In conventional molecular electronic structure calculations the one‐ and two‐electron integrals are computed over all space using basis functions that are typically Gaussians or Slater-type orbitals. For example, the one‐electron integrals have the general form
\begin{equation}\label{eq:one_electron}
h_{pq} = \int \phi_p^*(\mathbf{r}) h(\mathbf{r}, -i\hbar\nabla) \phi_q(\mathbf{r})  d\mathbf{r},
\end{equation}
and the two‐electron integrals are given by
\begin{equation}\label{eq:two_electron}
h_{pqrs} = \int\!\!\int \phi_p^*(\mathbf{r}_1)\phi_q^*(\mathbf{r}_2)\frac{1}{|\mathbf{r}_1-\mathbf{r}_2|}\phi_r(\mathbf{r}_1)\phi_s(\mathbf{r}_2) d\mathbf{r}_1 d\mathbf{r}_2.
\end{equation}
These integrals account for the full spatial extent of the electronic wave functions, ensuring that both short- and long-range interactions are included.

In the R--matrix inner-region problem one uses the same physical operators (kinetic energy, electron--nucleus attraction, and electron--electron repulsion), but solves them on a finite domain: all coordinate integrals are taken only over the inner region, typically a sphere of radius $a$ that encloses the target. In this sense, the differential and Coulomb operators are unchanged; what changes is the region of integration and the choice of basis functions and boundary treatment at $r=a$.

Accordingly, the one-electron integrals are evaluated as
\begin{equation}\label{eq:inner_one_electron}
h_{pq} = \int_{r\le a} \phi_p^*(\mathbf{r}) h(\mathbf{r}, -i\hbar\nabla) \phi_q(\mathbf{r})  d\mathbf{r},
\end{equation}
with basis functions adapted to the finite domain. In particular, continuum orbitals are constructed for the inner region and are allowed to have nonzero amplitude at the boundary; the boundary condition at $r=a$ is commonly enforced by augmenting the inner-region Hamiltonian with a surface (Bloch) operator, which restores Hermiticity on the truncated domain and provides the correct matching conditions \cite{blo57}. Long-range interactions beyond $r=a$ are then handled separately in the outer region.

Each (N+1)-electron eigenfunction $\Psi_k$ of the resulting Hamiltonian (associated with an energy $E_k$) can be expanded as:
\begin{equation}\label{eq:psi_expansion}
\Psi_k = \mathcal{A}\sum_{ij} a_{ijk} \Phi_i u_{ij} +
\sum_\alpha b_{\alpha k} \chi_\alpha.
\end{equation}
Where $\mathcal{A}$ is the antisymmetrisation operator which enforces antisymmetry of the $(N+1)$-electron wavefunction under the exchange of any two electrons, $\Phi_i(x_1,\dots,x_N)$ are the N-electron target eigenstates (indexed by $i$), $u_{ij}(x_{N+1})$ are continuum orbitals for channel $(i,j)$, $a_{ijk}$ are the open-channel coefficients (with indices corresponding to target $i$, continuum $j$, and state $k$), $\chi_\alpha$ are purely bound ($L^2$) configurations, with $L^2$ coefficients $b_{\alpha k}$.

On the boundary $r = a$, only the continuum part contributes, defining the surface amplitudes. The R-matrix on the boundary is then given by the expansion:
\begin{equation}
R_{i j} = \frac{1}{2 a} \sum_{k} \frac{w_{i k}(a)w_{j k}(a)}{E_{k} - E}.
\end{equation}
Where the boundary amplitude $w_{i k}(a)$ is defined as:
\begin{equation}
w_{i k}(a) = \sum_{j} a_{i jk} u_{i j}(a),
\end{equation}
and the sum is taken over a given channel.

These equations encapsulate the influence of the short–range (inner) region on the scattering process and introduces the energy dependence of this process. In the outer region—where the wavefunction is determined by a simpler long–range potential—the R–matrix serves as the matching condition; in practice outer region solutions are usually obtained by
propagating this R-matrix \cite{jt474}.

A possible workflow is:
\begin{enumerate}[label=(\roman*)]
  \item A classical code (e.g.\ UKRmol+) computes the inner-region integrals, continuum orbitals, and boundary values $u_{ij}(a)$; a preceding quantum target-state routine provides the retained target eigenfunctions $\{\Phi_i\}$.
  \item The quantum device runs the variational inner-region algorithm for the $(N{+}1)$-electron problem and returns optimised angles $\{\theta^{*}_{ij}\}$, thereby fixing the mixing matrix $U(\boldsymbol{\theta}^*)$ and the channel coefficients $a_{ijk}$.
  \item A classical post-processing script evaluates the required matrix elements of $U(\boldsymbol{\theta}^*)$ (equivalently, translates $\boldsymbol{\theta}^*$ into $a_{ijk}$), then combines $a_{ijk}$ with the known boundary values $u_{ij}(a)$ to obtain the surface amplitudes $w_{ik}(a)$ and hence the $R$-matrix.
  \item A classical outer-region solver propagates the $R$-matrix (or equivalently uses the $w_{ik}(a)$) to obtain scattering observables such as cross sections.
\end{enumerate}

A single solution obtained in the inner region gives rise to solutions for a broad range of scattering energies. This occurs because the inner region eigenfunctions, which incorporate all the complex short–range correlation effects, are used to define boundary conditions that must be met by many different asymptotic states. Furthermore, the process of solving the inner region problem can be computationally very demanding, as it involves the diagonalisation of matrices whose size increases rapidly with the level of detail needed to accurately describe electron correlation. 

There are number of key differences between the R-matrix Hamiltonian described above and the standard molecular Hamiltonian solved by standard quantum chemistry problems. As already mentioned, it only considers a finite region and an important assumption of the R-matrix
method is that the target wavefunctions have zero amplitude outside this region. Conversely the continuum functions do have amplitude on the boundary but cannot in total contain more than a single electron. In quantum‐chemistry VQE one typically targets only the ground state; by contrast, R-matrix scattering requires many eigenstates in the given symmetry sector, including numerous excited pseudostates or continuum-type states \cite{jt332}. This makes a naïve one-eigenstate-at-a-time approach costly. The method
developed below is designed to accommodate these changes.

\section{Method}

We map the second-quantised Hamiltonian for the inner region problem to a qubit Hamiltonian through the standard Jordan--Wigner transformation, to obtain a qubit Hamiltonian suitable for quantum computation \cite{Whitfield2011, Kassal2011}, although in further work we plan to employ the more compact encodings we have developed \cite{Picozzi2023}.

Here and throughout Sec.~III, \(H\) denotes the {inner-region} Hermitian operator \(H_{\mathrm{in}}(a)=H+\mathcal{L}(a)\) (Fig.~\ref{fig:rmatrix_partition}), i.e.\ the Hamiltonian built from finite-domain integrals together with the Bloch surface operator that restores Hermiticity on \(r\le a\).

As with other VQA-based approaches, the practical performance of our method depends on optimisation landscape properties and on the susceptibility to vanishing gradients.
Barren plateaus can arise for certain ans\"atze/cost choices and can make training intractable as problem size grows \cite{McClean2018BarrenPlateaus}.
Moreover, the relationship between provable trainability guarantees and classical simulability is subtle, and understanding it is important when assessing where quantum resources provide an advantage \cite{Cerezo2025AbsenceBP}.
In this work we mitigate these issues pragmatically by using a structured Givens-rotation network with explicit symmetry constraints and by employing sequential subspace optimisation with a simple $\langle H\rangle$ objective, though a systematic scaling study of trainability for larger active spaces remains for future work.

\subsection{Number projection operator}

In the inner region formulation the Hamiltonian must be projected onto a subspace in which at most one electron occupies the continuum orbitals. In practice we define a projection operator $P$ such that
\begin{equation}
    P = P_{t,N+1}P_{c,0} + P_{t,N}P_{c,1}.
\label{eq:projection_operator}
\end{equation}
where $P_{t,n}$ projects onto configurations with $n$ electrons in the target orbitals and $P_{c,m}$ projects onto configurations with $m$ electrons in the continuum. Thus $P_{t,N+1}P_{c,0}$ enforces all $N+1$ electrons bound and none in the continuum, while $P_{t,N}P_{c,1}$ enforces one electron in the continuum and $N$ bound. 

In the Jordan–Wigner mapping, $P_{t,N}$ can be expressed as a polynomial of Pauli-$Z$ operators on the target-orbital qubits that enforces Hamming weight $N$. Detailed expressions for the qubit representation of the number projection operators have been provided by Picozzi \cite{picozzi2024pascal}.

Our ansatz enforces the single–continuum–occupancy constraint structurally, so the prepared state obeys
\begin{equation}
\ket{\psi(\boldsymbol{\theta})} = P\ket{\psi(\boldsymbol{\theta})},
\end{equation}
and plain energies can be evaluated as
\begin{equation}
\bra{\psi(\boldsymbol{\theta})} H \ket{\psi(\boldsymbol{\theta})}
\;=\;
\bra{\psi(\boldsymbol{\theta})} P H P \ket{\psi(\boldsymbol{\theta})}.
\end{equation}
However, for second–moment–based costs (i.e. variance, folded–Hamiltonian) one needs the expectation of the square of the projected Hamiltonian $\bra{\psi(\boldsymbol{\theta})} (P H P)^2 \ket{\psi(\boldsymbol{\theta})}$.

On any state with $\ket{\psi(\boldsymbol{\theta})} = P\ket{\psi(\boldsymbol{\theta})}$ this second moment can be evaluated as
\begin{align}
\bra{\psi(\boldsymbol{\theta})} (P H P)^2 \ket{\psi(\boldsymbol{\theta})}
&=
\bra{\psi(\boldsymbol{\theta})} P H P H P \ket{\psi(\boldsymbol{\theta})}\\
&=
\bra{\psi(\boldsymbol{\theta})} H P H \ket{\psi(\boldsymbol{\theta})},
\label{eq:second_moment_identity}
\end{align}
Where, in general, $\bra{\psi}PH^{2}P\ket{\psi}\neq\bra{\psi}HPH\ket{\psi}$.
For the qubit form of the number–projection operators we follow Ref.~\cite{picozzi2024pascal}. We see that this form can be useful in cases where one is not able to enforce the constraint from the quantum circuit (for a fault-tolerant example, in phase estimation).

\subsection{Quantum circuits for the $N+1$ electron problem}

In standard R-matrix calculations the inner-region step amounts to solving the eigenproblem of \(H_{\mathrm{in}}(a)\) in a basis consisting of (i) antisymmetrised channel functions \(\mathcal{A}\{\Phi_i u_{ij}\}\) and (ii) additional \(L^2\) configurations \(\{\chi_\alpha\}\), yielding eigenpairs \(\{E_k,\Psi_k\}\) and the expansion coefficients \(a_{ijk},b_{\alpha k}\) in Eq.~(\ref{eq:psi_expansion}). Our quantum formulation represents this same truncated basis on three qubit registers with total Hilbert space factorisation $\mathcal H=\mathcal H_{\rm sys}\otimes\mathcal H_{\rm sel}$, where ${\rm sys}:=t{+}c$ and ${\rm sel}:=a$. We first obtain the retained $N$-electron target states $\{\Phi_i\}$ from a separate quantum target-state calculation on the target register. We then build the $(N{+}1)$-electron circuit with $n=t+c+a$ qubits and a fixed state-preparation block $V$ that maps computational branches to the truncated channel/$L^2$ basis. In the open-channel sector of Eq.~(\ref{eq:psi_expansion}), the selector basis state $\ket{i}$ identifies the retained target state $\Phi_i$ and the continuum register encodes the coupled orbital index $j$ (or more generally a superposition over $j\in\mathcal{J}_i$), so the amplitudes of these branches are the coefficients $a_{ijk}$; for bound/$L^2$ branches, the selector label together with the occupations stored on the system register identifies the configuration $\chi_\alpha$ and its amplitude $b_{\alpha k}$. The variational block $U(\boldsymbol{\theta})\in SO(k)$ acts only within this $k$-dimensional trial space, and at convergence its amplitudes determine the coefficients entering Eq.~(\ref{eq:psi_expansion}); the boundary amplitudes then follow from $w_{ik}(a)=\sum_j a_{ijk}\,u_{ij}(a)$, with $u_{ij}(a)$ supplied classically.

The register definitions and index conventions are summarised in Table~\ref{tab:register_map}.

\begin{figure*}[htbp]
   \centering
   \includegraphics[width=0.6\textwidth]{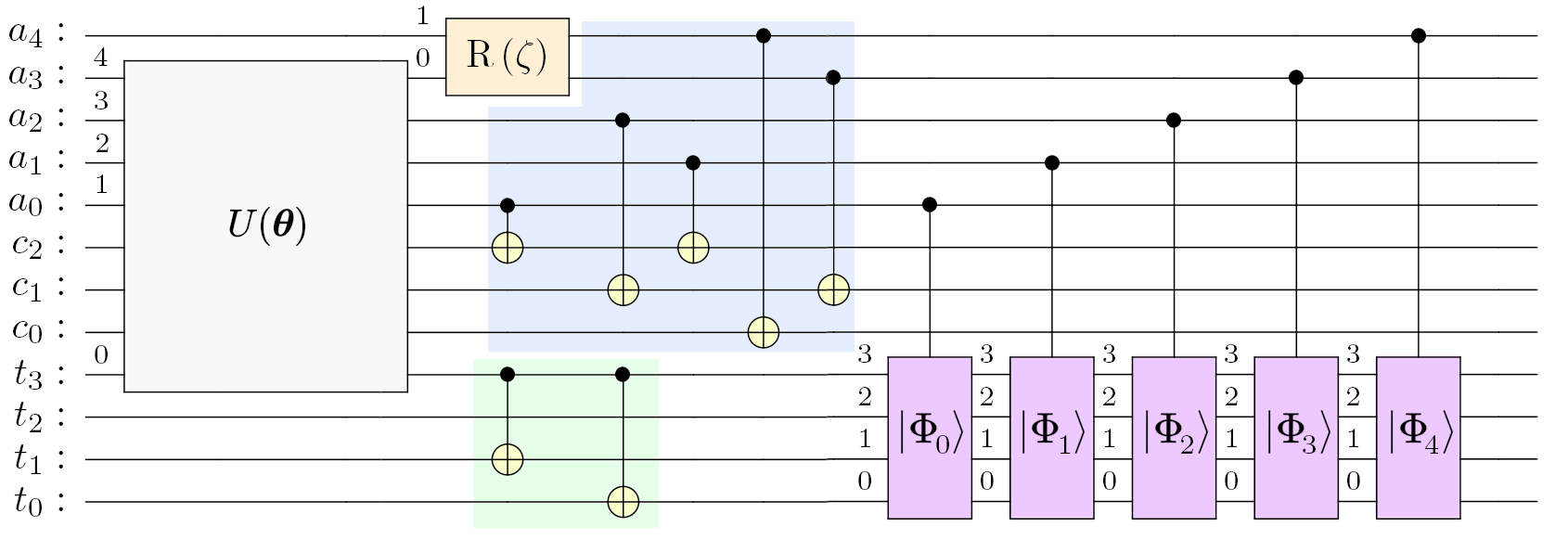}
  \caption{Schematic of the simultaneous eigenstate circuit showing the target orbital ($t$), continuum ($c$) and target-state selector ($a$) registers for the symmetry subspace with $S=\tfrac{1}{2}$, $M=-\tfrac{1}{2}$ and the irreducible representation  $B_{1u}$ of the point-group D$_{2h}$ used in the calculations. Some of the qubits that correspond to continuum spin-orbitals do not appear in the circuit due to symmetry.}
   \label{fig:simultaneous_circuit}
\end{figure*}

Our formalism also allows a binary selector encoding, in which the selector uses $\lceil\log_2 a\rceil$ qubits instead of $a$; this reduces qubit count but typically increases multi-controlled gate overhead. We therefore report resources below for the one-hot encoding used in our numerical experiments. Our formalism naturally allows for the inclusion of virtual orbitals (in the R-matrix sense of extra orbitals to complement the continuum description), with a modified form of the projection operator in Eq.~(\ref{eq:projection_operator}). A schematic of the circuit is shown in Figure~\ref{fig:simultaneous_circuit}. 

\begin{table}[htbp]
  \caption{Register-to-index map for the circuit representation of Eq.~(\ref{eq:psi_expansion}).}
  \centering
  \begin{tabular}{p{0.18\columnwidth} p{0.18\columnwidth} p{0.52\columnwidth}}
    \hline
    Register & Qubits & Role in Eq.~(\ref{eq:psi_expansion}) \\
    \hline
    Target ($t$) & $t$ & Encodes occupations of the target spin-orbitals; in the open-channel sector it stores the retained $N$-electron target state $\Phi_i$. \\
    Continuum ($c$) & $c$ & Encodes occupations of continuum spin-orbitals and the orbital index $j$ coupled to target channel $i$. \\
    Selector ($a$) & $a$ & Labels computational branches by $i\in\{0,\dots,a-1\}$. In the one-hot encoding, $\ket{i}$ is the single-excitation state with only the $i$th selector qubit equal to $\ket{1}$; for open channels it labels $\Phi_i$, while for bound/$L^2$ branches it is completed by the occupations on the system register to identify $\chi_\alpha$. \\
    \hline
  \end{tabular}
  \label{tab:register_map}
\end{table}

In the close--coupling approximation used in our H$_2$ demonstration, each target state $i$ couples to a single continuum orbital which we denote by a map $j(i)\in\{0,\dots,c-1\}$; more generally, each $i$ may couple to a set $\mathcal{J}_i$ of continuum orbitals and the circuit replaces a single CNOT by a controlled rotation that prepares a superposition over $j\in\mathcal{J}_i$.

\begin{figure*}[htbp]
    \centering
    \includegraphics[width=\textwidth]{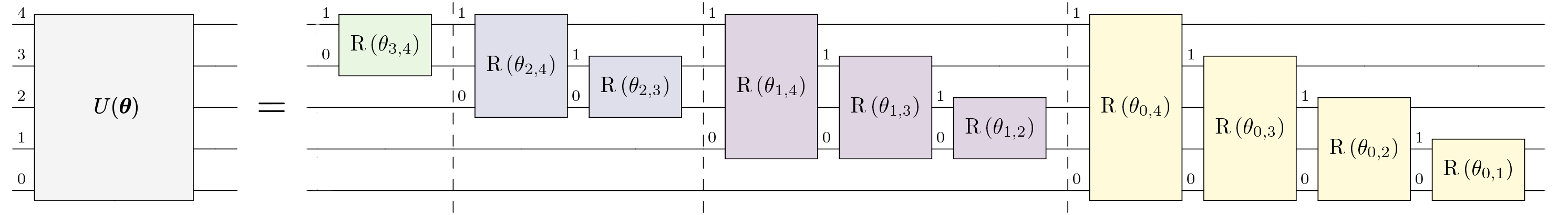}
    \caption{Schematic of the SO(5) rotation gate decomposition into a sequence of $\frac{5\times4}{2} = 10$ two-qubit Givens rotations with subspace optimisation.}
    \label{fig:so5_decomposition}
\end{figure*}

The circuit decomposes as follows:
\begin{enumerate}
    \item A parametric mixing gate \(U(\boldsymbol{\theta})\) that acts {only within the selected \(k\)-dimensional trial-state manifold} (implemented as a Givens-rotation cascade, Fig.~\ref{fig:so5_decomposition}), followed by the Clebsch--Gordan rotations \(R(\zeta)\) that are fixed by spin symmetry (in our 
    example, as in Figure~\ref{fig:simultaneous_circuit}, a single rotation 
    corresponding to the triplet state). The angle $\zeta$ is given by Eq.~\eqref{eq:clebsch-gordan} and the gate is the two-qubit Givens rotation of Eq.~\eqref{eq:givens_rotation} (the spin-symmetry Givens rotation derived from Eqs.~\eqref{eq:upper_multiplet}--\eqref{eq:clebsch-gordan}). This fixed CG block prepares the spin-adapted linear combinations of target state and incoming-electron spin that define the channel functions $\mathcal A\{\Phi_i u_{ij}\}$ entering Eq.~(\ref{eq:psi_expansion}); the subsequent variational block $U(\boldsymbol{\theta})$ then mixes those channels within the trial subspace.
    \item The target–continuum couplings (blue block in Figure~\ref{fig:simultaneous_circuit}, 
    CNOT gates with control on the target-state selector register and target on the 
    continuum register). Each selector qubit acts as control on one continuum 
    orbital qubit to ensure that, if a particular target eigenfunction is selected, 
    exactly one continuum orbital is excited in the correct symmetry channel. In the 
    close‐coupling approximation used in our example, each target eigenfunction couples to a 
    single continuum orbital; more generally, one can replace the CNOT by 
    parameterised controlled rotation to mix multiple continuum orbitals.
    \item The bound‐states gate (green block in Figure~\ref{fig:simultaneous_circuit}, CNOT gates 
    with control and targets on the target‐orbital register). A gate that entangles the 
    selector qubits with the target-orbital qubits, so that if the circuit has 
    selected a particular bound configuration, the correct pattern of target-orbital 
    occupations is realised. In simple examples, one can use CNOTs for each occupied 
    spin-orbital; for larger active spaces, this becomes a cascade of controlled Givens 
    rotations to generate the active‐space correlations.
    \item The sequence of controlled selector gates, each controlled on one of 
    the selector qubits and with targets on the target-orbital qubits.
\end{enumerate}

A central element of our quantum circuits is the one-parameter two-qubit rotation operation which mixes the $\ket{01}$ and the $\ket{10}$ states, implemented through a sequence of controlled-NOT (CNOT) gates interleaved with single-qubit rotations. Its matrix form is given by
\begin{equation}
R(\theta)=
\begin{pmatrix}
1 & 0 & 0 & 0\\[1mm]
0 & \cos \theta & \sin \theta & 0\\[1mm]
0 & -\sin \theta & \cos \theta & 0\\[1mm]
0 & 0 & 0 & 1
\end{pmatrix},
    \label{eq:givens_rotation}
\end{equation}
and it serves as the basic building block for establishing couplings in the circuit \cite{Kassal2011, Berry2018}.

In practical R-matrix calculations one enlarges the trial space by increasing the number of target states, continuum spin-orbitals, and $L^2$ configurations included in the truncation of Eq.~(\ref{eq:psi_expansion}). Convergence is reached by enlarging this space until the inner-region eigenvalues and boundary amplitudes entering the R-matrix sum are stable over the energy window of interest.

For the minimal H$_2$ demonstration (two target spin-orbitals, two continuum spin-orbitals and three selector qubits), the unoptimised CNOT count is $217$ and the circuit depth is $314$. In total, $7$ qubits are used, with $10$ variational parameters in the $SO(5)$ block, and $92$ Pauli terms are measured under the subspace protocol (see Table \ref{tab:resources_h2}).

\begin{table}[htbp]
  \caption{Circuit resource estimates for the minimal H$_2$ example.
We use $n=t{+}c{+}a=7$ qubits in the one-hot selector encoding: $t=2$ target spin–orbital qubits, $c=2$ continuum spin–orbital qubits, and $a=3$ selector qubits in the circuit implementation. Together with the occupations on the system register, these selector branches generate the five trial states listed in Sec.~IV. 
In a binary selector encoding the selector would require $\lceil\log_2 a\rceil$ qubits instead, at the expense of additional multi-controlled operations.}
  \centering
  \begin{tabular}{lr}
    \hline
    Number of qubits                  & 7              \\ 
    CNOT gates (unoptimised)                         & 217            \\ 
    Circuit depth                                  & 314            \\ 
    Variational parameters & 10             \\ 
    Pauli terms measured (subspace method)           & 92             \\ 
    \hline
  \end{tabular}
  \label{tab:resources_h2}
\end{table}

For fixed orbital bases, the total qubit count scales as $n=t+c+a$ in the one-hot selector encoding (or $n=t+c+\lceil\log_2 a\rceil$ in a binary selector encoding).
The variational mixing block introduced above is implemented as a fixed cascade of $\tfrac{k(k-1)}{2}$ two-qubit Givens rotations (Fig.~\ref{fig:so5_decomposition}), so the number of entangling blocks and the circuit depth of this block scale as $\mathcal{O}(k^2)$.
The target--continuum coupling layer scales as $\mathcal{O}(a)$ in the restricted close--coupling case (one continuum orbital per target state, i.e.\ one controlled operation per $i$), and more generally as $\mathcal{O}\!\bigl(\sum_i |\mathcal{J}_i|\bigr)$ when each channel couples to a set $\mathcal{J}_i$ of continuum orbitals via controlled rotations.
The number of distinct Pauli expectation values required to estimate $\langle H\rangle$ is set by the number of Pauli terms in the qubit Hamiltonian after symmetry reduction; crucially, our SSO protocol avoids squared-Hamiltonian observables and requires only $\langle H\rangle$ measurements in each sequential optimisation round.

Once the optimized parameters $\boldsymbol{\theta}^*$ have been found, the open-channel coefficients $a_{ijk}$ in Eq.~(\ref{eq:psi_expansion}) are obtained directly from the implemented variational unitary $U(\boldsymbol{\theta}^*)$ restricted to the $k$-dimensional trial-state (channel) subspace.
Concretely, we work in the single-excitation (one-hot) basis $\{\ket{i}\}$ on the selector register, so $\ket{i}$ labels the retained target channel index $i$.
In the restricted close-coupling case used in our H$_2$ demonstration, each channel $i$ is associated with a single continuum orbital $j(i)$, and the coefficient multiplying $\Phi_i u_{i j(i)}$ in $\Psi_k$ is the amplitude of the corresponding basis state after applying $U(\boldsymbol{\theta}^*)$.
Equivalently,
\begin{equation}
    a_{i\,j(i)\,k} = U_{i k}(\boldsymbol{\theta}^*),
\end{equation}
where $U_{ik}(\boldsymbol{\theta}^*)\equiv \bra{i}U(\boldsymbol{\theta}^*)\ket{k}$ is the $(i,k)$ matrix element of the implemented $SO(k)$ rotation in the trial-state basis. This identification is specific to the restricted close-coupling case with one continuum orbital per channel.
No additional variational optimisation is required to obtain $a_{ijk}$: once $\boldsymbol{\theta}^*$ is known, the entries $U_{ik}(\boldsymbol{\theta}^*)$ are fixed by the same Givens network used in the circuit and can be evaluated classically.
Optionally, $|a_{i\,j(i)\,k}|^2$ can be obtained directly as the probability of the outcome $(\text{selector}=i,\ \text{continuum}=j(i))$ when measuring $\ket{\Psi_k}$, and relative signs/phases can be determined by a two-outcome interference measurement within a $\{\ket{i},\ket{i'}\}$ subspace.
More generally, if channel $i$ couples to multiple continuum orbitals $j\in\mathcal{J}_i$, the circuit replaces the single controlled excitation by a controlled rotation that prepares a superposition over $j$, and the corresponding set of coefficients $\{a_{ijk}\}_{j\in\mathcal{J}_i}$ are obtained from the amplitudes in the joint selector--continuum single-excitation manifold, rather than from a single matrix element $U_{ik}(\boldsymbol{\theta}^*)$. For larger molecules and larger trial spaces, workflow (iii) is still classical: one evaluates the same fixed sequence of Givens rotations defining $U(\boldsymbol{\theta}^*)$, extracts the required amplitudes $a_{ijk}$ in the trial subspace, and then forms $w_{ik}(a)=\sum_j a_{ijk}u_{ij}(a)$. If one reconstructs the full trial-space rotation explicitly, this post-processing scales as $\mathcal{O}(k^3)$, with an additional dependence on $\sum_i |\mathcal{J}_i|$ in the general multi-continuum case, rather than requiring an additional quantum optimisation.

In the simultaneous optimisation algorithm, this mixing gate uses $\tfrac{k(k-1)}{2}$ angles (and only $k-1$ in the single-eigenstate case). The trial states include both bound configurations ($N + 1$ electrons in target orbitals) and open-channel functions ($N$ electrons in target orbitals plus one in the continuum). In our construction, only states with a single selector-qubit excitation (and similarly only one continuum excitation) are considered, which simplifies the gate structure.

\subsection{Spin symmetry}
In the R-matrix formalism, we couple the target of spin $S_t$ to an incoming electron of spin $s=\tfrac12$ to  a total spin $S$ with projection $M$.  
There are four combinations for the target state spin quantum numbers that can in principle contribute, namely the combinations with $S_t = S \pm \tfrac12$ and $M_t = M \pm \tfrac12$.
Whilst the coupling between the  upper multiplet with
$S_t = S - \tfrac12$ and the lower multiplet with
$S_t = S + \tfrac12$ is dynamical, the couplings between states with $M_t = M - \tfrac{1}{2}$ and $M_t = M + \tfrac{1}{2}$ for total spin in the upper multiplet and the lower multiplet are each fixed by symmetry. Each spin multiplet admits a two‑term expansion in terms of the half‑angle $\zeta$:
\begin{equation}
\bigl|S,M\bigr\rangle = \cos\zeta
\bigl|S_t,M-\tfrac12\bigr\rangle\otimes\bigl|\tfrac12, +\tfrac12\bigr\rangle
+\\
\sin\zeta
\bigl|S_t,M+\tfrac12\bigr\rangle\otimes\bigl|\tfrac12, -\tfrac12\bigr\rangle,
\label{eq:upper_multiplet}
\end{equation}
Where $\zeta$ is determined by the Clebsch--Gordan relations for the coupling between the target eigenfunctions and the incoming electron as:
\begin{equation}
\zeta = \begin{cases}
\frac12\arccos\!\Bigl(\frac{M}{S}\Bigr), & \text{for } S_t = S-\tfrac{1}{2},\\[1mm]
\frac{\pi}{2}
+\frac12\arccos\!\Bigl(\frac{M}{S+1}\Bigr), & \text{for } S_t = S+\tfrac{1}{2}.
\end{cases}
\label{eq:clebsch-gordan}
\end{equation}
Substituting Eq.~(\ref{eq:upper_multiplet}) into the open-channel sector of Eq.~(\ref{eq:psi_expansion}), each spin-adapted channel function with total quantum numbers $(S,M)$ takes the form
\begin{equation}
\mathcal{A}\{\Phi_i u_{ij}\}\big|_{S,M}
=\mathcal{A}\Bigl\{
  \bigl[\cos\zeta_i\;\Phi_i^{(M_t=M-\frac{1}{2})}\,u_{ij}^{\uparrow}
  +\sin\zeta_i\;\Phi_i^{(M_t=M+\frac{1}{2})}\,u_{ij}^{\downarrow}\bigr]
\Bigr\},
\label{eq:cg_bridge}
\end{equation}
where the superscripts on $\Phi_i$ and $u_{ij}$ indicate the target spin projection $M_t$ and the continuum-electron spin projection $m_s$, respectively, and $\zeta_i$ is fixed by the target spin $S_t$ via Eq.~(\ref{eq:clebsch-gordan}). In the quantum circuit this spin coupling is implemented by the fixed Givens rotation $R(\zeta_i)$ of Eq.~(\ref{eq:givens_rotation}), acting on the pair of qubits that encode $M_t$ and $m_s$ for channel $i$.

Using fixed Clebsch--Gordan rotations to enforce total–spin symmetry has known quantum–circuit realizations~\cite{BaconChuangHarrow2006}.
Our contribution is to apply a CG–enforcing circuit within a variational scattering ansatz, to our knowledge for the first time in the scattering literature.

\subsection{Cost functions}
A critical element of our variational formulation is the choice of cost function and each optimisation strategy. We consider several alternatives (where the cost functions are defined on the system register, see below for the treatment of the target-state selector register).

After Jordan--Wigner mapping, we write the system Hamiltonian as a Pauli expansion
$H=\sum_{\mu} h_{\mu} P_{\mu}$ with $P_{\mu}\in\{I,X,Y,Z\}^{\otimes (t+c)}$.
Expectation values such as $\langle H\rangle$ are estimated by measuring the Pauli strings $P_{\mu}$ and classically combining outcomes with weights $h_{\mu}$.
The number--projection operator $P$ in Eq.~(\ref{eq:projection_operator}) is a polynomial in Pauli-$Z$ operators on the target/continuum registers \cite{picozzi2024pascal}, so the products $HPH$ and $(HPH)$-based moments can likewise be evaluated via Pauli decomposition of the composite observable:
we expand $HPH$ into Pauli strings on the same registers and estimate $\langle HPH\rangle$ by measuring those strings (grouped into commuting sets in practice).
All numerical results reported in Sec.~IV use this explicit Pauli-decomposition evaluation of $\langle H\rangle$ and $\langle HPH\rangle$ on a noiseless simulator.

We denote by $\{\ket{\psi_\mu}\}_{\mu=0}^{k-1}$ the $k$ orthonormal $(N{+}1)$-electron {trial basis states} used in the simultaneous protocols (listed explicitly in Sec.~IV).
Given a shared mixing unitary $U(\boldsymbol{\theta})\in SO(k)$ acting within the span of these trial states, we define
\begin{equation}
\ket{\psi_\mu(\boldsymbol{\theta})} \equiv U(\boldsymbol{\theta})\ket{\psi_\mu}.
\end{equation}
In our implementation, the full state lives on the target-orbital register ($t$), continuum register ($c$), and the target-state selector register ($a$), but the Hamiltonian $H$ acts non-trivially only on the system register ${\rm sys}=t{+}c$, i.e.\ as $H\otimes I_{\rm sel}$ with ${\rm sel}=a$.
Here the index $\mu$ labels trial/output states in the $k$-dimensional simultaneous subspace and is distinct from the channel/target-state label $i$ used in Eq.~(\ref{eq:psi_expansion}).

\begin{enumerate}
    \item \textbf{Single state with variance cost function}\\
    In the single eigenstate approach the cost function is defined as the variance of the Hamiltonian with respect to the trial state,
    \begin{equation}
    C(\boldsymbol{\theta})= \langle H P H \rangle_{\ket{\psi(\boldsymbol{\theta})}} - \langle H \rangle_{\ket{\psi(\boldsymbol{\theta})}}^2.
    \end{equation}
    Where $H^2$ has been replaced by $H P H$ to enforce correct occupancy. This sometimes fails to return the full eigenspectrum.
    
    \item \textbf{Single state with folded Hamiltonian cost function}\\
    An alternative approach is to first evaluate the energy expectation for each trial state: the approximate energy $\Tilde{E}$ is first evaluated for the trial state, and then used to construct a Hamiltonian \cite{WangZunger1994, TaziThom2024} where, as long as the objective state eigenvalue is the closest eigenvalue to the energy of the trial state $\Tilde{E}$:
    \begin{equation}
    C(\boldsymbol{\theta})= \langle HPH\rangle_{\ket{\psi(\boldsymbol{\theta})}} - 2\Tilde{E} \langle H \rangle_{\ket{\psi(\boldsymbol{\theta})}} + \Tilde{E}^2 
    \end{equation}
    Here again the projection operator $P$ is used to enforce the occupancy constraints.
    
    \item \textbf{Simultaneous optimisation with sum of variances}\\
    In the simultaneous optimisation of all five eigenstates (see explicit list in Results section), the cost function is constructed as the sum of the variances of the individual eigenstates,
    \begin{equation}
C(\boldsymbol{\theta})= \sum_{\mu=0}^{k-1}\left[\langle HPH \rangle_{\ket{\psi_\mu(\boldsymbol{\theta})}} - \langle H \rangle_{\ket{\psi_\mu(\boldsymbol{\theta})}}^2\right].    \end{equation}
    This approach, while necessitating separate evaluations for each eigenstate, recovers the full spectrum of that eigensector within $10^{-7}\,E_h$.
    
    \item \textbf{Simultaneous optimisation with subspace optimisation and simple Hamiltonian expectation value}\\
    The most computational resource efficient method is based solely on the Hamiltonian expectation value,
    \begin{equation}
    C_\mu(\theta_{\mu, \mu + 1}, \dots, \theta_{\mu, k - 1})= \langle H \rangle_{\ket{\psi_\mu(\boldsymbol{\theta})}}.
    \end{equation}
    In this approach the unitary is designed to enforce a subspace structure that naturally preserves orthogonality without extra penalty terms. The overall unitary is constructed by decomposing an SO(5) rotation  into a sequence of ten two-qubit Givens rotations each parameterized by an angle. The trial states are first ordered by their energy. The gate decomposition begins by coupling the second highest eigenstate with the highest; subsequently, the third highest eigenstate is coupled to the two highest and so forth. Initially, all rotation parameters are set to zero to give the identity operator. The optimisation is carried out in reversed order, in that the coupling parameters that connect the lowest trial state (in our example circuit $\theta_{0, 1}, \theta_{0, 2}, \theta_{0, 3}$ and $\theta_{0, 4}$) to the other trial states are optimised in the first variational round, whilst the remaining parameters are left fixed: this effectively fixes the ground state subspace and restricts the remaining 4 parameters to a parametrization of $SO(4)$. Once the lowest eigenstate is determined, the same procedure is repeated for the couplings between the second lowest trial state and the higher states (namely $\theta_{1, 2}, \theta_{1, 3}$ and $\theta_{1, 4}$), each time automatically constraining the evolution to a smaller subspace that remains orthogonal to the lower eigenstates. This sequential subspace optimisation bypasses the need for higher-order terms such as $H^2$ and the explicit projection operator, and  corresponds to a diagonalisation of the Hamiltonian $H$, where at each variational round a $k$-dimensional block is block diagonalised into a single eigenvalue and a $k-1$-dimensional block.  Orthogonality is guaranteed by construction: at every step the ansatz is a single shared $SO(k)$ rotation $U(\boldsymbol{\theta})$ acting on an orthonormal trial basis, so the prepared set $\{\ket{\psi_\mu(\boldsymbol{\theta})}\}$ is orthonormal for any parameter values.   In the sequential subspace procedure, once the parameters that determine the lowest state are optimised, they are held fixed; the remaining free parameters then act only within the orthogonal complement subspace (equivalently, they parameterise the residual $SO(k-1)$ block), so subsequent rounds cannot reintroduce overlap with already-fixed lower states. A schematic of the decomposition of the SO(5) rotation gate $U(\boldsymbol{\theta})$ is given in Figure~\ref{fig:so5_decomposition}.
    SSVQE~\cite{Nakanishi2019} and MCVQE~\cite{Parrish2019} optimise a shared ansatz over a target subspace; orthogonality is handled via state-design or cost shaping, often requiring overlap measurements or explicit penalty terms. In contrast, our SSO protocol fixes an $SO(k)$ Givens cascade that sequentially isolates eigenstates by block-diagonalising $H$ without penalties or Gram–Schmidt. This makes the objective purely $\langle H\rangle$ (without squared terms), and pairs naturally with our coherent-summation readout to cut measurement count.

\end{enumerate}
For the algorithms 2., 3. and 4. all eigenvalues are recovered within $10^{-7}\,E_h$.

\subsection{Coherent summation protocol}

We introduce a coherent summation readout on the target--state selector register which gives an alternative route to estimate the expectations of the operators of interest without uncomputing the selector qubits. 

Let the joint state of the selector register (with $a$ qubits) and the system register (target--orbital and continuum--orbital qubits) be
\begin{equation}
\lvert \Psi\rangle =\sum_{\ell=0}^{2^a-1}\lvert \ell\rangle_{\rm sel}\otimes\lvert\varphi_\ell\rangle_{\rm sys},
\end{equation}
where $\lvert \ell\rangle_{\rm sel}$ are computational-basis states and $\lvert\varphi_\ell\rangle_{\rm sys}$ are the corresponding conditional system states. The goal is to access the coherently summed (generally unnormalised) system state
\begin{equation}
\lvert \psi_{u}\rangle_{\rm sys}\;\equiv\;\big(\langle u\rvert_{\rm sel}\otimes I_{\rm sys}\big)\lvert \Psi\rangle
=\frac{1}{2^{a/2}}\sum_{\ell=0}^{2^a-1}\lvert \varphi_\ell\rangle_{\rm sys},
\label{eq:psi_u_def}
\end{equation}
where we define the uniform superposition state on the selector in terms of the Hadamard gate $\mathrm{H}$ (not to be confused with the Hamiltonian $H$) as
\begin{equation}
\lvert u\rangle_{\rm sel} \equiv \mathrm{H}^{\otimes a}\lvert 0\cdots 0\rangle_{\rm sel}.
\end{equation}
Equivalently,
\begin{equation}
\lvert \psi_{u}\rangle_{\rm sys}
=\big(\langle 0\cdots 0\rvert_{\rm sel} \mathrm{H}^{\otimes a}\otimes I_{\rm sys}\big)\lvert \Psi\rangle.
\label{eq:psi_u_hadamard}
\end{equation}
We are interested in expectation values of $H$ on the corresponding normalised postselected state.

To implement this projection using only computational-basis measurements, we apply a Hadamard gate to each selector qubit and then measure the selector in the computational basis. Postselecting the outcome $0\cdots0$ after the Hadamard gates prepares the normalised version of $\ket{\psi_u}$ on the system register. In practice we:
\begin{enumerate}[label=(\roman*)]
\item apply $\mathrm{H}^{\otimes a}$ on the selector register;
\item measure the selector register in the computational basis and postselect the outcome $0\cdots0$; and
\item on the postselected shots, measure the Pauli strings required to estimate $\langle H\rangle$ on the system register.
\end{enumerate}
Let $p_0$ be the probability of obtaining $0\cdots0$ after the Hadamard gates and let $\langle H\rangle_{0}$ denote the conditional expectation value of $H$ on the system given this outcome. Then
\begin{equation}
\langle H\rangle_{0}
=\frac{\langle \psi_{u}\lvert H\rvert \psi_{u}\rangle}{\langle \psi_{u}\mid \psi_{u}\rangle},
\qquad
p_0=\langle \psi_{u}\mid \psi_{u}\rangle,
\label{eq:postselect_relations}
\end{equation}
so that $\langle \psi_{u}\lvert H\rvert \psi_{u}\rangle = p_0\,\langle H\rangle_{0}$ can be reconstructed if desired.

This provides an alternative to coherently uncomputing (erasing) the selector register, which in our setting would require reversing the full selector-controlled target preparation and channel-dependent target--continuum couplings; such compute--uncompute cleanup typically increases the depth and entangling-gate count of the relevant state-preparation portion by a factor of roughly $1.5$--$2\times$ \cite{Bennett1973,Bennett1989}.

The postselection probability can be written as
\begin{align}
p_0 &=\big\langle \Psi \big| \Big(\mathrm{H}^{\otimes a}\lvert 0\cdots0\rangle\!\langle 0\cdots0\rvert \mathrm{H}^{\otimes a}\Big)_{\rm sel}\otimes I_{\rm sys} \big| \Psi \big\rangle
\\&=\big\langle 0\cdots 0 \big| \mathrm{H}^{\otimes a}\rho_{\rm sel} \mathrm{H}^{\otimes a}\big| 0\cdots 0 \big\rangle,
\end{align}
where $\rho_{\rm sel}=\mathrm{Tr}_{\rm sys}\!\left(\lvert \Psi\rangle\!\langle \Psi\rvert\right)$ is the reduced state of the selector register prior to the Hadamard gates. In the one-hot encoding used throughout this work, the selector support is restricted to the Hamming-weight-one subspace spanned by $\{\ket{e_i}\}_{i=1}^{a}$, which implies the general bound
\begin{equation}
p_0 \le \frac{a}{2^{a}}.
\label{eq:p0_onehot_bound}
\end{equation}
By contrast, for a binary selector encoding with $m=\lceil \log_2 a\rceil$ qubits (so $2^{m}\approx a$), amplitudes spread comparably over the encoded channels can yield $p_0=O(1)$, leading to more favourable shot scaling for this postselected readout.

\section{Results}

\begin{figure*}[htbp]
  \centering
  \begin{minipage}[t]{0.48\textwidth}
    \centering
    \includegraphics[width=\linewidth]{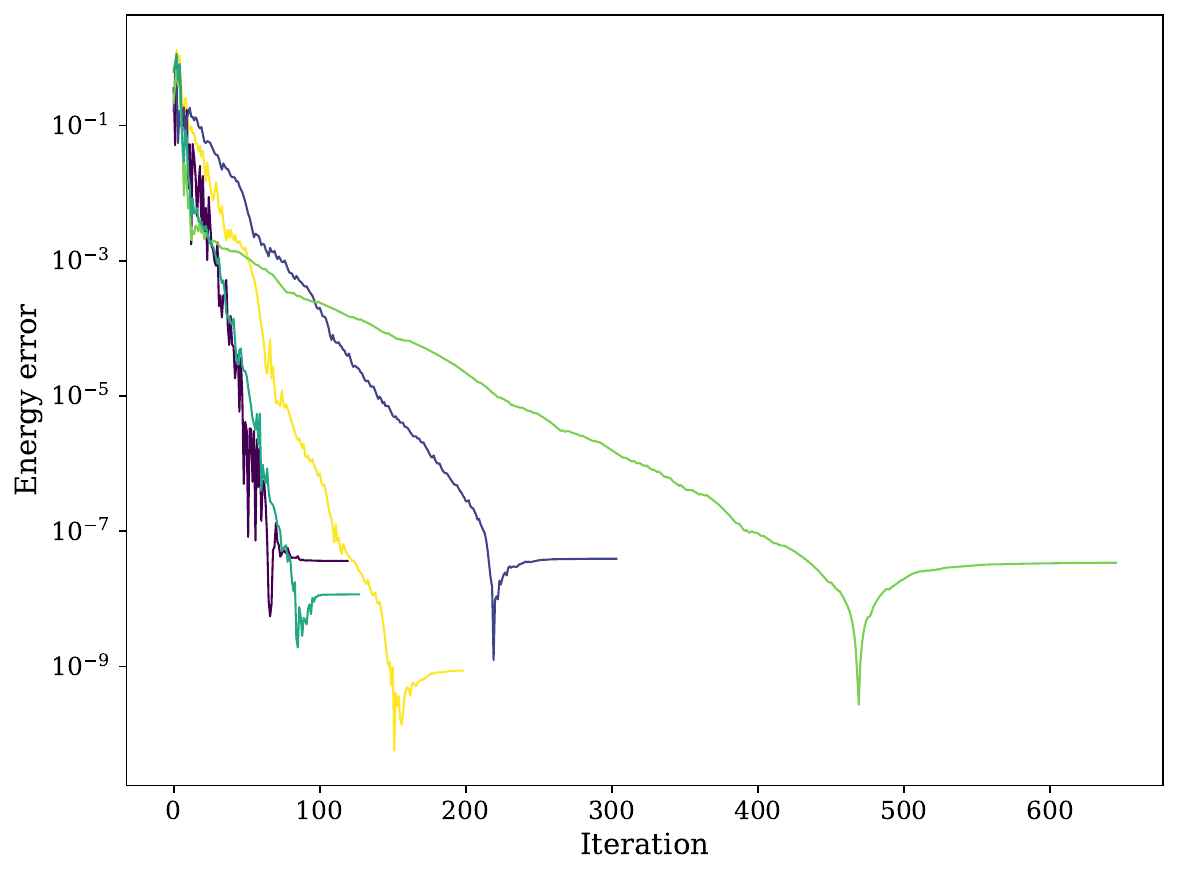}
  \end{minipage}\hfill
  \begin{minipage}[t]{0.48\textwidth}
    \centering
    \includegraphics[width=\linewidth]{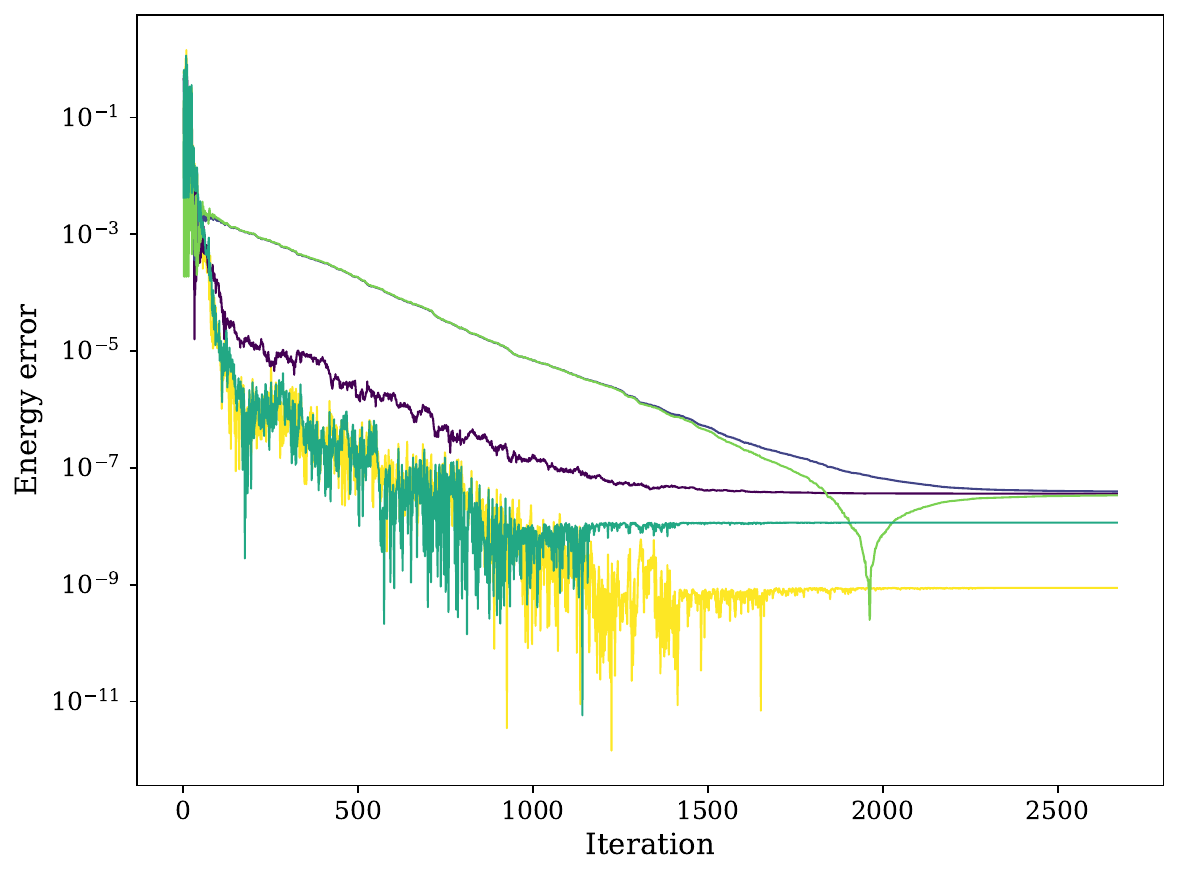}
  \end{minipage}

  \vspace{0.8ex}

  \begin{minipage}[t]{0.48\textwidth}
    \centering
    \includegraphics[width=\linewidth]{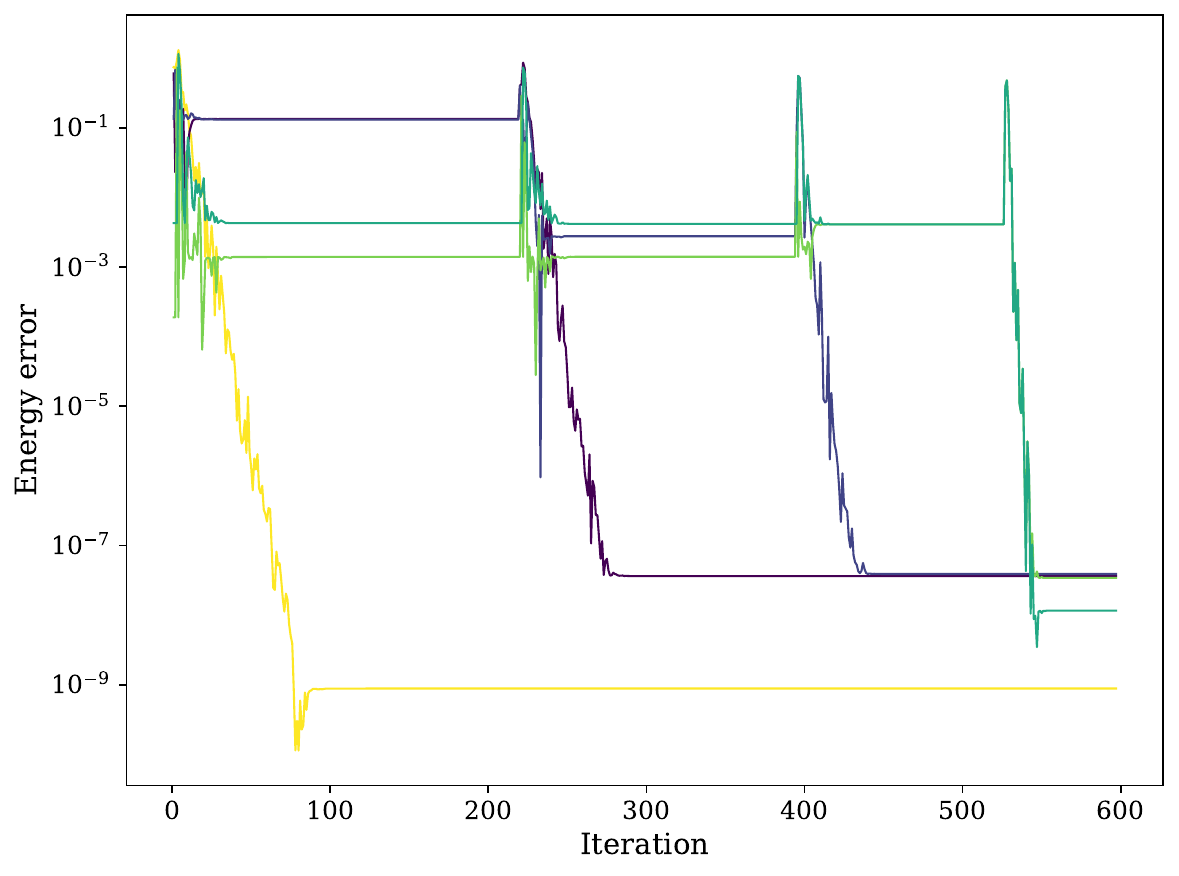}
  \end{minipage}\hfill
  \begin{minipage}[t]{0.48\textwidth}
    \centering
    \includegraphics[width=\linewidth]{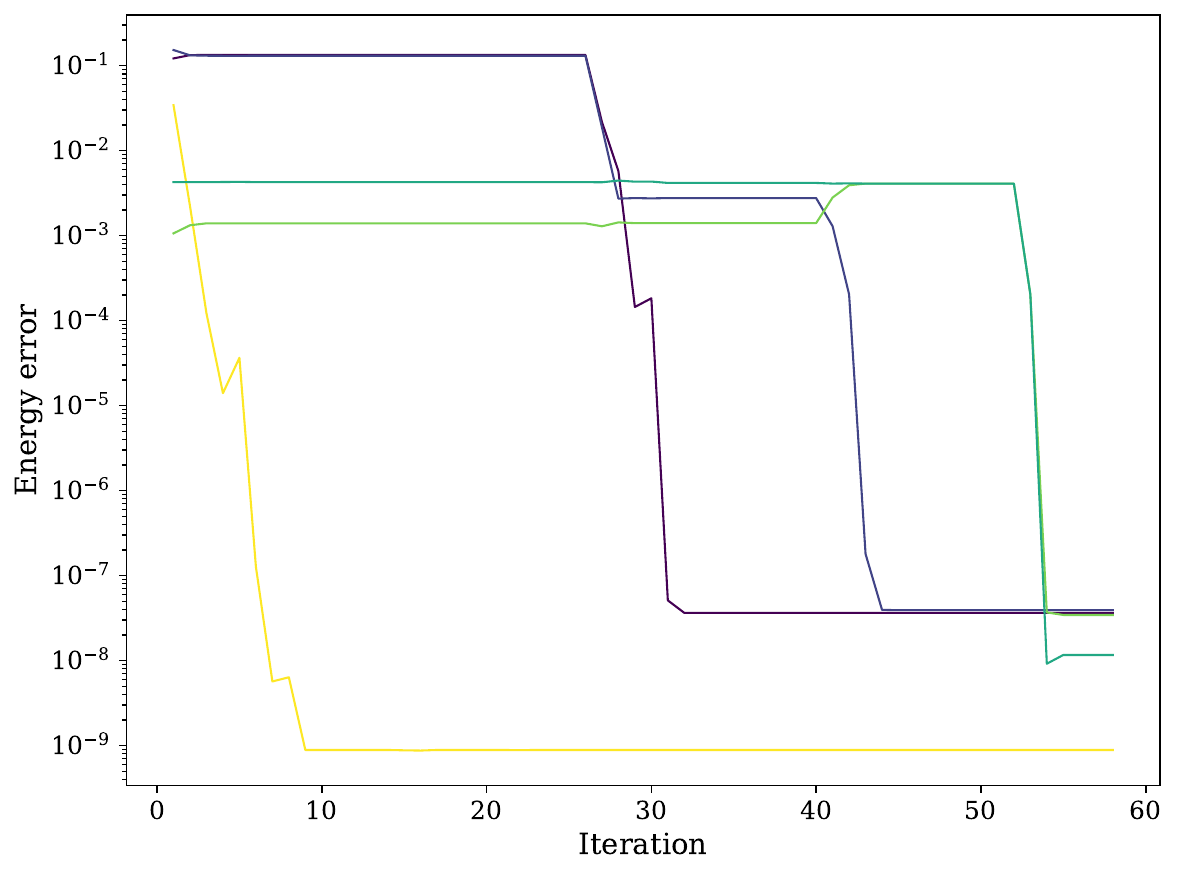}
  \end{minipage}

\caption{%
  Comparison of convergence behaviours for different cost functions.
  Here the energy error for state $k$ at a given optimisation step is
  $\Delta E_k \equiv \bigl|E_k^{(\mathrm{var})}-E_k^{(\mathrm{ref})}\bigr|$ (in Hartree),
  where $E_k^{(\mathrm{ref})}$ is the corresponding exact eigenvalue obtained by classical diagonalisation
  of the same $(N{+}1)$-electron qubit Hamiltonian in the chosen symmetry sector.
  The \(N+1\) eigenstates are labelled (in order of increasing energy) as: \\
    \(\textcolor{c1}{\blacksquare}\): \(\ket{1,\tfrac12,-\tfrac12,\mathrm{B_{1u}}}\)
    \(\textcolor{c2}{\blacksquare}\): \(\ket{2,\tfrac12,-\tfrac12,\mathrm{B_{1u}}}\) 
    \(\textcolor{c3}{\blacksquare}\): \(\ket{3,\tfrac12,-\tfrac12,\mathrm{B_{1u}}}\)
    \(\textcolor{c4}{\blacksquare}\): \(\ket{4,\tfrac12,-\tfrac12,\mathrm{B_{1u}}}\)
    \(\textcolor{c5}{\blacksquare}\): \(\ket{5,\tfrac12,-\tfrac12,\mathrm{B_{1u}}}\)\\
  Top left panel: Convergence diagram for single state optimisation using the folded Hamiltonian cost function; \\
  Top right panel: Convergence behaviour for the simultaneous eigenstate optimisation when minimising the sum of variances;\\
  Bottom left panel: Convergence diagram for simultaneous optimisation with subspace optimisation and simple Hamiltonian expectation value using COBYLA;\\
  Bottom right panel: Convergence diagram for sequential subspace optimisation using SLSQP with the same initial angles and stopping criterion as the COBYLA run.
  All curves correspond to a single optimisation instance (no averaging over multiple random initialisations). 
  All variational angles are initialised to zero (so that the initial mixing block is the identity), and energies are evaluated without finite-shot sampling; the non-monotonic features therefore reflect optimiser updates and landscape structure rather than measurement noise.
}
  \label{fig:convergence_comparison}
\end{figure*}

To illustrate our approach we consider the example of electron scattering from the hydrogen molecule. In this model problem the (bound) target orbitals are taken from a minimal basis as $\sigma_{g}$ and $\sigma_{u}$ (these orbitals are of symmetry a$_g$ and b$_{1u}$ in the D$_{2h}$ point group used in the calculations and their double occupancy gives rise to states of $A_g$ and $B_{1u}$ symmetry) while the two continuum orbitals are chosen so that one of them is of $a_{g}$ and the other one is of $b_{1u}$ symmetry. Orbitals and integrals
were obtained from a standard R-matrix calculation using a minimal (STO-3G) basis for H$_2$, a Gaussian Type orbital (GTO) representation of the continuum and an R-matrix radius of $r = 10\,a_0$ using the
UKRmol+ code \cite{Masin2020}.

We first solve the full $N$-electron target problem with $N = 2$ using the separate quantum target-state routine described in Sec.~III.B, and thereby obtain the five target eigenfunctions allowed by symmetry:
\begin{align*}
\ket{\Phi_0} &= c_{0}\ket{\sigma_g^2} -c_{1}\ket{\sigma_u^2},\\
\ket{\Phi_1} &= c_{0}\ket{\sigma_u^2} +c_{1}\ket{\sigma_g^2},\\
\ket{\Phi_2} &= \frac{1}{\sqrt{2}}\Bigl(\ket{\sigma_u^\uparrow\sigma_g^\downarrow} -\ket{\sigma_u^\downarrow\sigma_g^\uparrow}\Bigr),\\
\ket{\Phi_3} &= \frac{1}{\sqrt{2}}\Bigl(\ket{\sigma_u^\uparrow\sigma_g^\downarrow} +\ket{\sigma_u^\downarrow\sigma_g^\uparrow}\Bigr),\\
\ket{\Phi_4} &= \ket{\sigma_u^\downarrow\sigma_g^\downarrow}.
\end{align*}
Where $c_{0} \approx 0.99361$ and $c_{1} \approx 0.11287$.
In our simulation, we consider the symmetry subspace with total spin quantum number $S=\tfrac{1}{2}$, projected quantum number $M=-\tfrac{1}{2}$ that lies in the point-group irreducible representation $B_{1u}$. Here the retained trial space has dimension $k=5$: four of these trial states are open-channel branches built from the target eigenfunctions above, and one is the bound three-electron configuration. In the circuit implementation this trial basis is generated jointly by the $a=3$ selector qubits and the occupations on the target/continuum registers, rather than by the selector register alone.
Our circuit prepares trial wavefunctions corresponding to the close-coupling trial states:
\begin{align*}
\ket{\psi_0} &= \ket{\sigma_g^2 \sigma_{u\downarrow}},\\
\ket{\psi_1} &= \ket{\Phi_0}\otimes\ket{1b_{1u \downarrow}},\\
\ket{\psi_2} &= \ket{\Phi_1}\otimes\ket{1b_{1u \downarrow}},\\
\ket{\psi_3} &= \ket{\Phi_2}\otimes\ket{1a_{g \downarrow}},\\
\ket{\psi_4} &= \sqrt{\frac{2}{3}}\ket{\Phi_3}\otimes\ket{1a_{g \downarrow}} - \sqrt{\frac{1}{3}}\ket{\Phi_4}\otimes\ket{1a_{g \uparrow}}.
\end{align*}

The trial states $\ket{\psi_1}$, $\ket{\psi_2}$, $\ket{\psi_3}$, and $\ket{\psi_4}$ represent configurations with two electrons in a target orbital and one electron in the continuum, whereas $\ket{\psi_0}$ corresponds to the only state allowed by symmetry with all three electrons in bound orbitals.
These states are each prepared by juxtaposing a one-qubit Pauli-X operator $X_{q_i}$ to the circuit with all variational parameters set to zero.

In the eigensector with total spin $S=\tfrac{1}{2}$ and projection $M=-\tfrac{1}{2}$, the singlet target channel ($S_t = S - \tfrac{1}{2} = 0$) is entirely given by the $\lvert S_t=0,M_t=0\rangle\otimes\lvert\tfrac{1}{2},-\tfrac{1}{2}\rangle$ component and no Clebsch--Gordan rotation is required. By contrast, for the triplet target channel ($S_t = S + \tfrac{1}{2} = 1$) the angle $\zeta$ defined by the Clebsch--Gordan relations (\ref{eq:clebsch-gordan}) fixes the mixing between the $\lvert S_t=1,M_t=0\rangle\otimes\lvert\tfrac{1}{2},-\tfrac{1}{2}\rangle$ and the $\lvert S_t=1,M_t=-1\rangle\otimes\lvert\tfrac{1}{2},+\tfrac{1}{2}\rangle$ components.

We perform the classical optimisation using the derivative-free Constrained Optimization BY Linear Approximations (COBYLA) algorithm \cite{Powell1994, Powell1998} on a noiseless simulation.  We consider a number of cost functions:

\begin{enumerate}
    \item \textbf{Single state with variance cost function}\\
    In some instances this approach is unable to recover all the eigenvalues; for this reason we do not use it for the main benchmarks reported in Fig.~\ref{fig:convergence_comparison} and focus on the three approaches below.    
    
    \item \textbf{Single state with folded Hamiltonian cost function}\\
    This approach required approximately 1,397 cost function evaluations to recover the five eigenstates, each corresponding to the evaluation of 1,080 Pauli terms. The results are shown in the upper panel of Figure~\ref{fig:convergence_comparison}.
    
    \item \textbf{Simultaneous optimisation with sum of variances}\\
    This method required 13,345 cost function evaluations for 1,080 Pauli terms. The results are shown in the middle panel of  Figure~\ref{fig:convergence_comparison}.
    
    \item \textbf{Simultaneous optimisation with subspace optimisation and simple Hamiltonian expectation value}\\
    This method required only 573 cost function evaluations involving 92 Pauli terms. The results are shown in lower panel of Figure~\ref{fig:convergence_comparison}.
\end{enumerate}

The distinct behaviours in Fig.~\ref{fig:convergence_comparison} reflect the optimisation protocol: in the folded-Hamiltonian case (top) each eigenstate is optimised independently (five separate runs whose convergence traces are superimposed), whereas in the sum-of-variances (middle) and sequential subspace optimisation (bottom) the parameters are optimised in a single simultaneous procedure for all five states.

The sharp dips in $\Delta E_k$ are consistent with the optimiser moving between attraction basins (and, in the SSO case, with stage-wise release/locking of different parameter subsets), leading to abrupt improvements in the targeted state’s energy estimate.

For the last three algorithms all the eigenvalues were recovered as the converged values within $10^{-7}\,E_h$, the residual discrepancy expected from the stopping criterion.

To assess sensitivity to the choice of classical optimiser, we repeated the sequential subspace optimisation (SSO) experiment using sequential least-squares quadratic programming (SLSQP) instead of COBYLA (with the same initial angles) and found convergence in 41 cost-function evaluations, with final energy errors of order $10^{-8}\,E_h$. The corresponding SSO convergence trace is shown in the bottom-right panel of Fig.~\ref{fig:convergence_comparison}.

\section{Discussion}
The algorithm we have introduced fulfils several critical objectives. First, because our ansatz enforces the single--continuum occupancy constraint by construction, we can evaluate energies directly as $\langle \psi \lvert H \rvert \psi \rangle$ without explicitly measuring a projected observable such as $PHP$ (or the $HPH$ moments that arise in variance/folded--Hamiltonian costs). In our minimal $\mathrm{H}_2$ instance, the plain qubit Hamiltonian requires measuring $92$ Pauli strings (Table~\ref{tab:resources_h2}), whereas the explicitly projected observable used in the moment-based costs expands to $1080$ Pauli strings (Sec.~IV), i.e.\ a reduction of $(1080-92)/1080 \approx 91.5\%$ (about $12\times$ fewer terms). The precise reduction is system-dependent: it is set by how many Pauli-$Z$ monomials appear in the projector for the chosen active space/constraint and by algebraic cancellations in the Pauli expansion of the composite observable. Second, the circuit depth is on the order of a few hundred layers for our minimal $\mathrm{H}_2$ instance (depth $314$ with $217$ CNOTs before any hardware-aware compilation; Table~\ref{tab:resources_h2}). While this two-qubit-gate count will introduce non-negligible error on present-day NISQ hardware, the depth remains moderate compared with approaches that require deep Trotterised time evolution or phase estimation, and it can be reduced further through compilation and hardware-efficient Givens implementations. Third, the optimal rotation angles in the Givens network correspond directly to the boundary amplitudes of the R-matrix, thereby providing immediate access to the quantities of interest in scattering theory \cite{jt474,Picozzi2023}.

In established R-matrix implementations, the dominant inner-region cost is the construction and diagonalisation (or iterative solution) of a large Hamiltonian in the truncated symmetry-adapted channel/$L^2$ basis of dimension $k$, which grows rapidly with the number of target orbitals, continuum orbitals, and $L^2$ configurations; direct diagonalisation scales as $\mathcal{O}(k^3)$ time and $\mathcal{O}(k^2)$ memory, while symmetry reduction and iterative eigensolvers mitigate (but do not remove) this bottleneck. Qualitatively, as the physical description is made more realistic (more target orbitals, more continuum orbitals, more channels, and more $L^2$ configurations), this truncated inner-region basis size $k$ typically grows combinatorially because it counts many-electron configurations in the chosen symmetry sector.
Our approach targets this bottleneck by replacing explicit matrix diagonalisation with state preparation on $n$ qubits and expectation-value estimation of a qubit Hamiltonian, while exploiting circuit structure to reduce measurement overhead: the SSO protocol enforces orthogonality by construction and avoids $H^2$-type observables, so that only $\langle H\rangle$ measurements are required per sequential round.
While we do not claim a general, proven asymptotic quantum advantage over the best classical iterative methods at present, the method provides a concrete route by which quantum hardware could become advantageous as the trial-space dimension increases: the quantum device represents trial states without storing the full $k\times k$ matrix, and multiple eigenstates required for scattering can be obtained via a fixed $SO(k)$ cascade whose parameter count and depth scale as $\mathcal{O}(k^2)$.

Among the four optimisation strategies examined, the simultaneous subspace approach stands out for its efficiency. By decomposing the SO($N$) rotation into a cascade of two‑qubit Givens operations and enforcing orthogonality through sequential subspace optimisation, we reduce both the variational parameter count and the measurement overhead. In our hydrogen‑molecule test case this method recovered all eigenvalues to $10^{-7}\,E_h$ using only 573 cost function evaluations and 92 Pauli terms, compared with over ten thousand evaluations for the sum of variance scheme.

Although our circuit still spans all \((N+1)\)-electron configurations (including the bound-state sector), its structure is such that the variational parameters determine the open-channel coefficients \(a_{ijk}\) directly, so one never needs to isolate or compute the bound-state amplitudes \(b_{\alpha k}\). Concretely, we couple each \(N\)-electron target eigenstate only to its corresponding continuum orbital(s), and arrange the \(SO(k)\) rotation so that the resulting matrix elements yield the desired \(a_{ijk}\). As a result, once the circuit is optimised, the coefficients \(a_{ijk}\) are obtained from the same implemented Givens network without any further variational postprocessing, even though the full set of bound configurations appears in the Hilbert space. In this way, we avoid both deep Trotter decompositions (as in time-dependent methods) and large ancilla registers (as in quantum-walk schemes), while keeping the depth on the order of a few hundred (depth $314$ for $\mathrm{H}_2$ in a minimal basis; Table~\ref{tab:resources_h2}) and using $O(t+c+a)$ qubits in the one-hot selector encoding (or $O(t+c+\lceil\log_2 a\rceil)$ with a binary selector).\\
Looking ahead, this framework opens several avenues for advancement. One must extend the number of continuum orbitals to obtain realistic results which may require the  incorporation of more sophisticated encodings to further compress the qubit resource requirements \cite{Picozzi2023}. Error mitigation techniques and hardware‑efficient realisations of Givens rotations will be essential to translate these algorithms to larger systems in the form of both larger targets and more sophisticated representations of the target. The direct correspondence between circuit parameters and R-matrix boundary data opens the door to the integration of quantum processors into established scattering codes, offering a promising hybrid quantum classical paradigm for high‑precision collision physics. Finally, our architecture shows how a {CG symmetry–enforcing circuit} can be integrated into variational scattering ans\"atze, and it should transfer to other spin–coupled targets (atomic, nuclear, and model field–theory settings)~\cite{Sharma2024,Xing2023_KohnVQLS}.

We emphasize that, in contrast to prior quantum-computing studies of atom--molecule or nuclear scattering \cite{Xing2023_KohnVQLS,Sharma2024,Yusf2025} and atomic-model electron dynamics \cite{Chan2023}, the present work targets genuine electron--molecule scattering within the R-matrix formalism.

\section{Conclusion}

We have presented a variational quantum algorithm tailored to the inner‑region problem of electron--molecule scattering within the R-matrix approach using a variational framework \cite{Peruzzo2014, Aspuru2005}. Our approach employs a number‑projection operator to enforce physical occupancy constraints, a rotation‑based ansatz to generate the required symmetry‑adapted subspaces, and a coherent summation protocol to simplify observable estimation. Benchmarking on the hydrogen‑molecule system demonstrates that all scattering eigenvalues in a given symmetry sector are obtained while using a shallow circuit and minimal measurement resources. By making the open-channel coefficients, and hence the R-matrix boundary amplitudes, directly accessible from the optimised circuit, our method provides straightforward access to scattering observables. This work represents a significant step towards the practical application of quantum computers in molecular collision studies and lays the groundwork for future extensions to more complex targets and continuum representations.

\section*{Author contributions}
DP conceived and developed the central ideas, including the Hamiltonian formulation, circuit and algorithm design. VG provided the molecular integrals from the UK Molecular R-matrix Codes (UKRmol+) used to construct the Hamiltonian and contributed significantly to the discussion. JT and JDG offered valuable insights and discussions. All authors contributed to the manuscript.

\section*{Data availability}
The code that implements the variational algorithms described in this work, together with the molecular integrals and per-iteration energies needed to reproduce the four numerical experiments reported in Sec.~III, is openly available in a Zenodo-archived GitHub repository~\cite{picozzi_rmatrix_dataset_2026}.

\section*{Acknowledgments}
D.P.'s research is supported by an EPSRC Postdoctoral Prize Fellowship [EP/W524335/1] at UCL and an EPSRC Research Fellowship [EP/S021582/1] at the London Centre for Nanotechnology. D.P.’s PhD was supported by an EPSRC Industrial CASE studentship [EP/T517793/1].

\bibliography{paper}

\end{document}